 \documentclass[aps,superscriptaddress,showpacs,twocolumn,nofootinbib]{revtex4}

\usepackage{graphicx}% Include figure files
\usepackage{bm}% bold math
\usepackage{amsmath,epsfig}
\usepackage{tabularx}

\begin{document}

\title{Generation of hierarchically correlated multivariate symbolic sequences} 
%\subtitle{with an application to the assessment of bootstrap confidence in phylogenetic analysis}
%\titlerunning{Generation of hierarchically correlated multivariate symbolic sequences}
\author{Michele Tumminello}
\affiliation{Dipartimento di Fisica e Tecnologie Relative, 
Universit\`a di Palermo, Viale delle Scienze, I-90128 Palermo, Italy}
%\affiliation{CNR-INFM, Unit\`a di Palermo, Palermo, Italy}

\author{Fabrizio Lillo}
\affiliation{Dipartimento di Fisica e Tecnologie Relative, 
Universit\`a di Palermo, Viale delle Scienze, I-90128 Palermo, Italy}
\affiliation{Santa Fe Institute, 1399 Hyde Park Road, Santa Fe, NM 87501, U.S.A.}
%\affiliation{Istituto Nazionale di Fisica Nucleare, Sezione di Catania, Catania, Italy}

\author{Rosario N. Mantegna}
\affiliation{Dipartimento di Fisica e Tecnologie Relative, 
Universit\`a di Palermo, Viale delle Scienze, I-90128 Palermo, Italy}
%\affiliation{Istituto Nazionale di Fisica Nucleare, Sezione di Catania, Catania, Italy}

\date{\today}

\begin{abstract}
We introduce an algorithm to generate multivariate series of symbols from a finite alphabet with a given hierarchical structure of similarities. The target hierarchical structure of similarities is arbitrary, for instance the one obtained by some hierarchical clustering procedure as applied to an empirical matrix of Hamming distances. The algorithm can be interpreted as the finite alphabet equivalent of the recently introduced hierarchically nested factor model (M. Tumminello et al. EPL 78 (3) 30006 (2007)). The algorithm is based on a generating mechanism that is different from the one used in the mutation rate approach.  We apply the proposed methodology for investigating the relationship between the bootstrap value associated with a node of a phylogeny and the probability of finding that node in the true phylogeny.
\end{abstract}


\maketitle

\section{Introduction}

Symbolic sequences are investigated in many different fields, including information theory, biological sequence analysis, linguistics, chaotic time series, and communication theory. A lot of efforts have been devoted to devise algorithms for generating univariate or multivariate sequences with given statistical properties \cite{Li,Buldyrev,Makse,Makse2,Izrailev}.  Since pair correlations are often used to describe the dependence between variables, the problem of generating symbolic sequences with given pair correlation properties is of particular interest. Many algorithms have been proposed for generating symbolic sequences with given {\it univariate} correlation structure, e.g. given autocorrelation and to generate symbolic sequences with given {\it multivariate} correlation structure, e.g. given cross correlation among pair of sequences \cite{Emrich,Gange,Lee}. In this second case one wants to generate multivariate sequences of symbols according to some given properties of pair similarities. In this paper we propose an algorithm for generating multivariate sequences with a given similarity structure of {\it hierarchical} nature. This protocol is inspired by an algorithm recently introduced by us \cite{EPL2007} to generate hierarchically organized multivariate sequences with variables which are continuously distributed. 
The applications of the algorithm here proposed are manifold. For example, in phylogenetic analysis the characteristics of the investigated species are coded in discrete (symbolic) variables, such as nucleotides, amino acids, discrete characters, and phylogenetic algorithms give as an output a hierarchical tree. Our method gives the possibility of simulating the system without making any assumption on the evolutionary dynamics of the system.   

As a specific application of the generation algorithm, in this paper we consider a common problem in phylogenetic analysis, specifically the assessment through bootstrap analysis of the statistical confidence of a phylogenetic tree. Phylogeny is the study of evolutionary relations among different elements (for example, organisms or languages). There are many different algorithms to reconstruct a phylogenetic tree from a set of data. One of the key problems in phylogenetic analysis is the assessment of the  accuracy of a given tree feature (e.g. a node or an internal branch). Since a statistical theory of the errors of a phylogenetic method is usually difficult to achieve, a common approach to assess the accuracy of the features of a phylogenetic tree is bootstrap analysis \cite{Felsenstein}.  By sampling with replacements the data matrix and by applying the tree reconstruction algorithm to each bootstrap replica, one can obtain a confidence value of a feature by computing the fraction of replica trees that shares the considered feature with the original tree. In a seminal paper, Hillis and Bull \cite{Hillis} showed that this fraction is an underestimation of the probability of inferring the correct feature for bootstrap proportions larger than $40\%$. By using computer simulations of evolution dynamics of sequences they showed, for example, that ``bootstrap proportions of $\ge 70\%$ usually correspond to a probability of $\ge 95\%$ that the corresponding clade is real" \cite{Hillis}.  
The result of Hillis and Bull is based on a generic evolutionary model with a per-symbol constant mutation rate. While in molecular evolution this seems to be a natural starting model, in other contexts, such as language, culture or technology evolution, mutation rate and dynamical models based on it might be more vague concepts. Since our generation algorithm is independent of any dynamical assumption, we believe it may be well suited for application in these contexts. In this paper we apply our generation algorithm to the assessment of bootstrap confidence in phylogenetic analysis. We perform a simulation analysis similar to the one presented by Hillis and Bull in Ref.  \cite{Hillis} but using our generation algorithm.
Similarly to them we find that the bootstrap proportion underestimates the probability that a clade inferred from sample data belongs to the true phylogeny.

The paper is organized as follows. In Section \ref{algorithm} we present our algorithm for generating multivariate symbolic sequences with a given hierarchical similarity structure. In Section \ref{bootstrap} we present the application of the algorithm to the assessment of bootstrap proportion as a measure of confidence. Section \ref{conclusions} concludes.

\section{Algorithm for generating hierarchically organized multivariate symbolic sequences}\label{algorithm}
In this section, we introduce an algorithm allowing to simulate multivariate series of symbols from a finite alphabet. The objective is to generate symbolic sequences with a hierarchical structure of similarities between the elements of the system. This structure may correspond, for instance, to the one revealed by a hierarchical clustering procedure that has been applied to an empirical matrix of Hamming similarities. In this sense our protocol is the finite alphabet equivalent of the Hierarchically Nested Factor Model (HNFM) that we have introduced in ref. \cite{EPL2007}.

Let {\bf X} be a set of series of symbols from a finite alphabet $A=\{a_1,...,a_p\}$. We indicate the length of each series with $T$ and we assume that the number of series in the set is $N$. Let us arrange the data {\bf X} in such a way that each column of ${\bf X}$ corresponds to a specific series. According to the Hamming distance we define the similarity of elements $i$ and $j$ as
\begin{equation}\label{hammingsim}
s(i,j)=\frac{1}{T}\sum_{k=1}^{T} \delta(x_{k i},x_{k j}),
\end{equation}
where $\delta(x_{k i},x_{k j})=1$ if $x_{k i}=x_{k j}$ and 0 otherwise. It is easy to show the following properties of $s(i,j)$:
\begin{eqnarray}\label{hammprop}
s(i,i)=1\\
s(i,j)\le \frac{1}{T} \sum_{k=1}^{T} 1=1\\
s(i,j)\ge \frac{1}{T} \sum_{k=1}^{T} 0=0
\end{eqnarray}
These properties show that $s(i,j)$ assumes rational values in the closed interval $[0,1]$. Furthermore, it can be shown that $s(i,j)$ is the result of a scalar product. Indeed each symbol $a_i$ of the alphabet can be mapped into a vector of length $p$ with all the components equal to zero but the $i-th$ component being equal to 1. Any series ${\bf x_k}$ of length $T$ can therefore be mapped into a vector ${\bf \tilde{x}_k}$ of length $T \cdot p$ by substituting symbols in the series with the corresponding binary mapping. We can rewrite Eq. (\ref{hammingsim}) in terms of series ${\bf \tilde{x}_i}$ as
\begin{equation}\label{hammingsimscal}
s(i,j)=\frac{1}{T}\sum_{k=1}^{T p}\tilde{x}_{k i}\tilde{x}_{k j} =\frac{1}{T} \, {\bf \tilde{x}_{i}} \cdot {\bf \tilde{x}_{j}}.
\end{equation}
The properties described in Eq.s (\ref{hammprop}-\ref{hammingsimscal}) imply that the matrix ${\bf S}$ of similarities $s(i,j)$ can be interpreted as a correlation matrix, because (i) it is positive definite as the result of scalar product of Eq. (\ref{hammingsimscal}), (ii) its diagonal elements are equal to 1 and (iii) all the elements $s(i,j)$ assume values in the range $[0,1]$. The latter condition indicates that similarities are described only in terms of positive numbers according to the Hamming distance. By applying a hierarchical clustering procedure to the matrix ${\bf S}$ of elements $s(i,j)$ of Eq. (\ref{hammingsim}) one obtains a filtered similarity matrix ${\bf S}^<$ and a dendrogram \cite{Anderberg}. 
\begin{figure}
\resizebox{1\columnwidth}{!}{\includegraphics{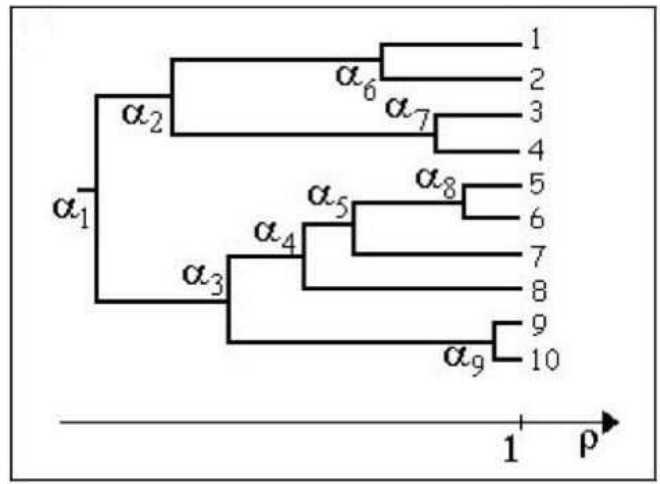}}
\caption{ Illustrative example of a rooted tree associated with a system of $N=10$ elements (leaves in the tree). The symbols $\{\alpha_1,...,\alpha_9\}$ labels the $N-1=9$ internal nodes.} 
\label{dendro}
\end{figure}
A dendrogram is a rooted tree, i.e. a tree in which a special node (the root) is singled out. This node is labeled $\alpha_1$ in the illustrative example of Fig.\ref{dendro}.  In the rooted tree, we distinguish between leaves and internal nodes. Specifically, vertices of degree $1$ represent leaves (vertices labeled $1, 2,..., 10$ in Fig. \ref{dendro}) while vertices of degree greater than 1 are internal nodes (vertices labeled $\alpha_1$, $\alpha_2$,..., $\alpha_9$ in Fig. \ref{dendro}). 
%
%We associate a \emph{genealogy} $G(i)$  ($G(\alpha_h)$) with each leaf $i$ (internal node $\alpha_h$).The genealogy is the ordered set of internal nodes connecting leaf $i$ (internal node $\alpha_h$) to the root $\alpha_1$. For instance, in Fig. \ref{dendro}, the genealogy associated with the leaf 3 is $G(3)=\{\alpha_7, \alpha_2, \alpha_1 \}$ and the genealogy of the internal node $\alpha_7$ is $G(\alpha_7)=\{\alpha_7, \alpha_2, \alpha_1 \}$. Note that the internal node $\alpha_7$ is included in $G(\alpha_7)$. Finally, 
We also say that an internal node $w$ is the {\it  parent} of the node $v$, and we use the notation $w=g(v)$, if $w$ immediately precedes $v$ on the path from the root to $v$. For example it is $\alpha_2=g(\alpha_7)$ in Fig. \ref{dendro}. Analogously we say that an internal node $w$ is a {\it son} of the node $v$ if $v$ is the parent of $w$, i.e. $v=g(w)$. In the example above $\alpha_7$ is the son of node $\alpha_2$.
Beside the topological structure, dendrograms obtained through standard hierarchical clustering algorithms applied to a matrix of Hamming similarities have also metric properties. In fact, clustering algorithms associate a similarity (correlation) coefficient  $\rho_{\alpha_i}$ with each internal node $\alpha_i$  \cite{Anderberg}. The whole information about the rooted tree is stored in the $N \times N$ matrix ${\bf S^<}$ of elements $s(i,j)=\rho_{\alpha_k}$, where $\alpha_k$ is the first internal node in which leaves $i$ and $j$ are merged together \cite{Anderberg}. For example, in Fig. \ref{dendro}, it is $s(3,7)=\rho_{\alpha_1}$ and $s(5,7)=\rho_{\alpha_5}$. Our internal node labeling implies that  $\rho_{\alpha_i} \le \rho_{\alpha_{i+1}}$.  
In ${\bf S^{<}}$ there are at most $N-1$ distinct elements. Exactly $N-1$ distinct elements are obtained in case of binary rooted trees. Since any rooted tree can be obtained from a rooted binary tree by introducing a degeneracy of nodes, in the following we consider binary rooted trees. 
The entries of ${\bf S}^<$ are non negative numbers as a consequence of dealing with the Hamming similarity. Therefore ${\bf S}^<$ is the correlation matrix of a suitable HNFM and as a consequence ${\bf S}^<$ is positive definite \cite{EPL2007}. In Ref. \cite{EPL2007} we have introduced an algorithm for generating continuously distributed variables having ${\bf S}^<$ as the correlation matrix. This is not the model we are looking for here because it cannot be used for simulations of symbols from a finite alphabet. Instead we are looking for a protocol allowing the generation of a set of series of symbols from the alphabet $A$, such that the similarity matrix of infinite length series generated by the protocol is exactly ${\bf S}^<$. \\

%For an infinite alphabet $A$ ($p\rightarrow \infty$) it turns out that $s(i,j)=P(i,j)=\rho_{\alpha_k}$ and therefore the proposed algorithm gives the required result. 

The algorithm we propose here generates one symbol at a time for all the leaves. The idea is to start from the root, generate a symbol and let this symbol propagate down the tree with some probability. If the symbol does not propagate one goes to the next node down the tree, generate a symbol and propagate it down the tree with some probability. The similarity between two leaves stems from the fact that a fraction of symbols was generated in a common ancestor of the two leaves. With finite alphabets however spurious similarities are observed. Let $P(i,j)$ denote the probability  that the symbol at node $i$ and at node $j$ has been generated in the same internal node. The expectation value of the similarity $s(i,j)$ is 
\begin{equation}\label{sijbias} 
E[s(i,j)]=P(i,j)+\frac{1-P(i,j)}{p}
\end{equation}
where the second term takes into account the fact that symbols in $i$ and $j$ can be equal despite the fact that they were generated in an independent way as a consequence of the finite dimension of the alphabet $A$. Therefore the first step of the algorithm consists in removing the bias due to the finiteness of the alphabet.

The algorithm works as follows.
\begin{enumerate}
\item{In order to remove the bias due to the finiteness of the alphabet, for each internal node one replaces\footnote{We observe that the transformation (\ref{newrho}) preserves the ranking of the correlation of nodes in the dendrogram. 
%Consider indeed two nodes $\alpha_k$ and $\alpha_h$ such that $\rho_{\alpha_k}>\rho_{\alpha_k}$. We have
%\begin{eqnarray}
%\rho^I_{\alpha_k}&= \rho_{\alpha_k}-\frac{1-\rho_{\alpha_k}}{p-1}=\rho_{\alpha_k} (1+\frac{1}{p-1})-\frac{1}{p-1}>\nonumber \\
%&>\rho_{\alpha_h} (1+\frac{1}{p-1})-\frac{1}{p-1}=\rho^I_{\alpha_h}.
%\end{eqnarray}
The ordering preservation implies that the topology of the dendrogram is not changed after the transformation.} $\rho_{\alpha_k}$, ($k=1,...,N-1$) with
\begin{equation}\label{newrho}
\rho^I_{\alpha_k}= \rho_{\alpha_k}-\frac{1-\rho_{\alpha_k}}{p-1}.  
\end{equation}
}
\item{One assigns a symbol $v_{\alpha_1}$ from the alphabet to the root node $\alpha_1$ of the dendrogram. A random number $u_1$, uniformly distributed in the interval $[0,1]$, is generated. If $u_1\le \rho^I_{\alpha_1}$ then the symbol $v_{\alpha_1}$ is assigned to all the elements of the system (leaves of the dendrogram) and to all the nodes $\alpha_i$ rooting at $\alpha_1$. In this case the assignment is complete and one goes to Step 5.}
\item{One moves to the nodes which are sons of $\alpha_1$ in the dendrogram. Moving along the branches of the dendrogram let us assume that we have reached a certain node $\alpha_k$. This implies that a symbol has still to be assigned to the leaves and the nodes rooting at $\alpha_k$. One randomly assigns a symbol $v_{\alpha_k}$ from the alphabet $A$ to the node $\alpha_k$. One then extracts a random number $u_k$. If 
\begin{equation}\label{condnode}
u_k \le \frac{\rho^I_{\alpha_k}-\rho^I_{g(\alpha_{k})}}{1-\rho^I_{g(\alpha_{k})}}  
\end{equation}
then one assigns the symbol $u_k$ to the leaves and the internal nodes rooting at $\alpha_k$, otherwise one moves to the next nodes (sons of $\alpha_k$). Once  a symbol has been assigned to the leaves and nodes belonging to a branch of the dendrogram then all of these nodes in the branch must be disregarded. }
\item{Once all nodes of the dendrogram have been explored (or disregarded because of the above condition) still some leaves could remain without an assigned symbol. One randomly assigns a symbol according to an uniform distribution to each of such leaves. }
\item{Consider the next symbol and go to Step 2.}
\end{enumerate}

By following this procedure, we have assigned a symbol to each leaf (element of the system) and to each internal node of the dendrogram. 
%and we can therefore move to simulate the next row of symbols by again randomly assigning a symbol to the root node of the dendrogram.
%
The validity of this algorithm in generating hierarchically organized sequences is based on the following

{\bf Proposition I}. {\it In the sequences generated according to the above algorithm, the probability $P(i,j)$ that the symbol at node $i$ and at node $j$ has been generated in the same internal node is $\rho^I_{\alpha_k}$, where $\alpha_k$ is the closest common ancestor (internal node) of  $i$ and $j$. }

The proof is given in the Appendix.  For a multivariate dataset generated according to the algorithm the expected value of the Hamming distance 
$s(i,j)$ between two leaves (elements) rooting first at node $\alpha_k$ is 
\begin{equation}\label{esij}
E[s(i,j)]=\rho^I_{\alpha_k}+\frac{1-\rho^I_{\alpha_k}}{p}=\rho_{\alpha_k}
\end{equation}
because of Proposition I and Eq.s (\ref{sijbias},\ref{newrho}).
%$P(i,j)$, which, according to the Proposition I, is equal to $\rho_{\alpha_k}$. 
%
Thus the generated dataset has similarity matrix which is on average equal to the similarity matrix  ${\bf S}^<$ of the dendrogram. The term on average has in this context two meanings. First, it means that for finite sequence length $T$ the similarity matrix averaged over many simulations is equal to ${\bf S}^<$. But it is also true that this equality holds also between ${\bf S}^<$ and one simulation of infinite length.

Our algorithm has some limitations.
First, in the current form the algorithm can be applied to trees where two leaves have the same similarity with their closest common ancestor. This is verified in many phylogenetic techniques, e.g. the unweighted pair group method using arithmetic averages (UPGMA)  \cite{Sokal}, but not in others, e.g. neighbor joining and maximum likelihood methods. We are currently developing extensions of the algorithm to the case when two leaves have different correlation with their closest common ancestor.
Second, the fact that  $\rho^I_{\alpha_k}$  is equal to the probability $P(i,j)$  implies that $\rho^I_{\alpha_k} \ge 0$, or, in other words, that $\rho_{\alpha_k} \ge \frac{1}{p}$ for any $k$. Therefore our method can be applied if all the $\rho_{\alpha_k}$ are larger or equal to $1/p$. This constraint 
indicates the impossibility of generating series of symbols with a correlation smaller than the 
correlation between independent random series with our method. We note that the same impossibility exists when one uses the mutation rate 
approach. Finally, when continuously distributed variables are considered ($p\rightarrow\infty$), we have obtained 
that the HNFM  \cite{EPL2007} can be defined if $\rho_{\alpha_k} \ge 0$ for any $k$, in agreement with what has 
been observed here. This facts suggest that the above constraint should be more related to the hierarchical 
organization of the system than to the specific method used to generate hierarchically organized data series.

\section{Test of bootstrapping as a method for assessing confidence in phylogenetic analysis}\label{bootstrap}

\begin{figure}
%\begin{center}
\includegraphics[width=0.48\textwidth]{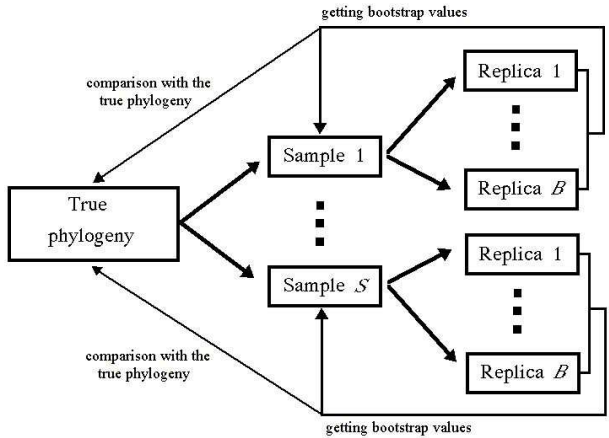}
\caption{Scheme of the procedure used to investigate the relationship between the bootstrap value associated with a node of a phylogeny based on sample data and the probability of finding that node in the true phylogeny.}
\label{Scheme}
%\end{center}
\end{figure}

%
%\begin{figure}
%\includegraphics[width=0.48\textwidth]{schemino}
%\includegraphics[width=0.45\textwidth]{Fig1}
%\caption{Scheme of the procedure used to investigate the relationship between the bootstrap value associated with a node of a phylogeny based on sample data and the probability of finding that node in the true phylogeny.} 
%\label{Scheme}
%\end{figure}
%

As an application of our generation algorithm in this section we investigate the relationship between the bootstrap value associated with a node of a phylogeny based on sample data and the probability of finding that node in the true phylogeny. The fact that the bootstrap value of a node is not equal to the probability that the node was present in the true phylogeny is known since the work of Hillis and Bull \cite{Hillis}. They simulated an evolutionary process of a set of sequences under a constant mutation rate and found that large bootstrap proportions typically underestimate the probability that the node is present in the true phylogeny used to simulate the data. Their result might be dependent on the evolutionary process they used in the simulation scheme. Here we want to adopt a similar testing procedure for the bootstrap by using the simulation algorithm introduced in Section \ref{algorithm}.

\begin{figure}
\resizebox{1\columnwidth}{!}{\includegraphics{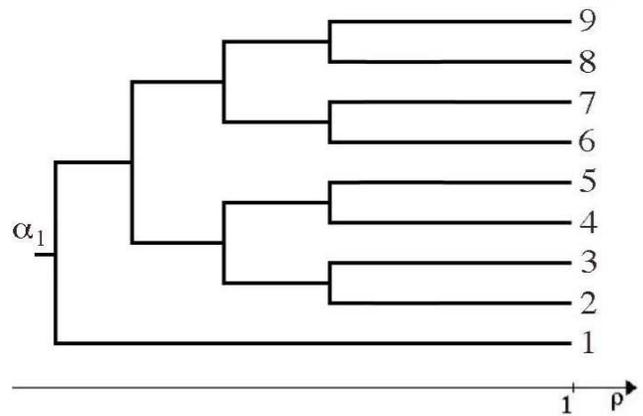}}
\caption{Phylogeny A of a system of $N=9$ elements (leaves in the tree). The symbol $\alpha_1$ labels the root node.} 
\label{DendroA}
\end{figure}

\begin{figure}
\resizebox{1\columnwidth}{!}{\includegraphics{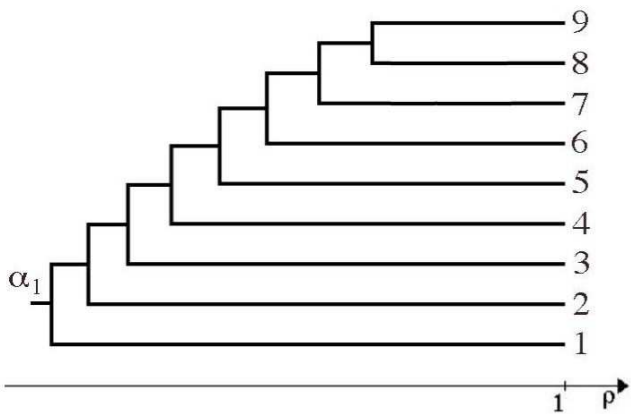}}
\caption{Phylogeny B of a system of $N=9$ elements (leaves in the tree). The symbol $\alpha_1$ labels the root node.} 
\label{DendroB}
\end{figure}

To this end, we choose the metric and topological properties of a phylogeny and we perform $S$ simulations according to the model described in the previous section. For the present application the dimension of the alphabet is 4, in order to simulate nucleic acids. We then extract the phylogeny associated with each simulation by using the Average Linkage Cluster Analysis \cite{Anderberg}, also known as UPGMA \cite{Sokal} and we estimate the accuracy of nodes (clades) in these simulations via the bootstrap technique \cite{Efron}. Once a bootstrap value has been associated with each node of each simulation, we count  the total number $n_{bt}$ of nodes in all the simulations having associated a bootstrap value in the range $[bt-5 \%, bt+5\%[$ with $bt=\{5\%,$ $15\%,...,95\%\}$ (the bootstrap value $100\%$ is included in the last interval). Then we measure the percentage of these $n_{bt}$ nodes that belong to the true phylogeny. Such percentage can be interpreted as the probability that a node with a bootstrap value belonging to the range $[bt-5 \%, bt+5\%[$ corresponds to a correct clade. This approach is also illustrated in Fig. \ref{Scheme}. Our simulations are based on two different dendrogram topologies, i.e. two different phylogenies. Specifically, we consider two of the topologies analyzed in ref. \cite{Hillis}. These topologies are shown in Fig. \ref{DendroA} and Fig. \ref{DendroB}.


\begin{figure} 
\begin{center}
              \resizebox{0.8\columnwidth}{!}{\includegraphics{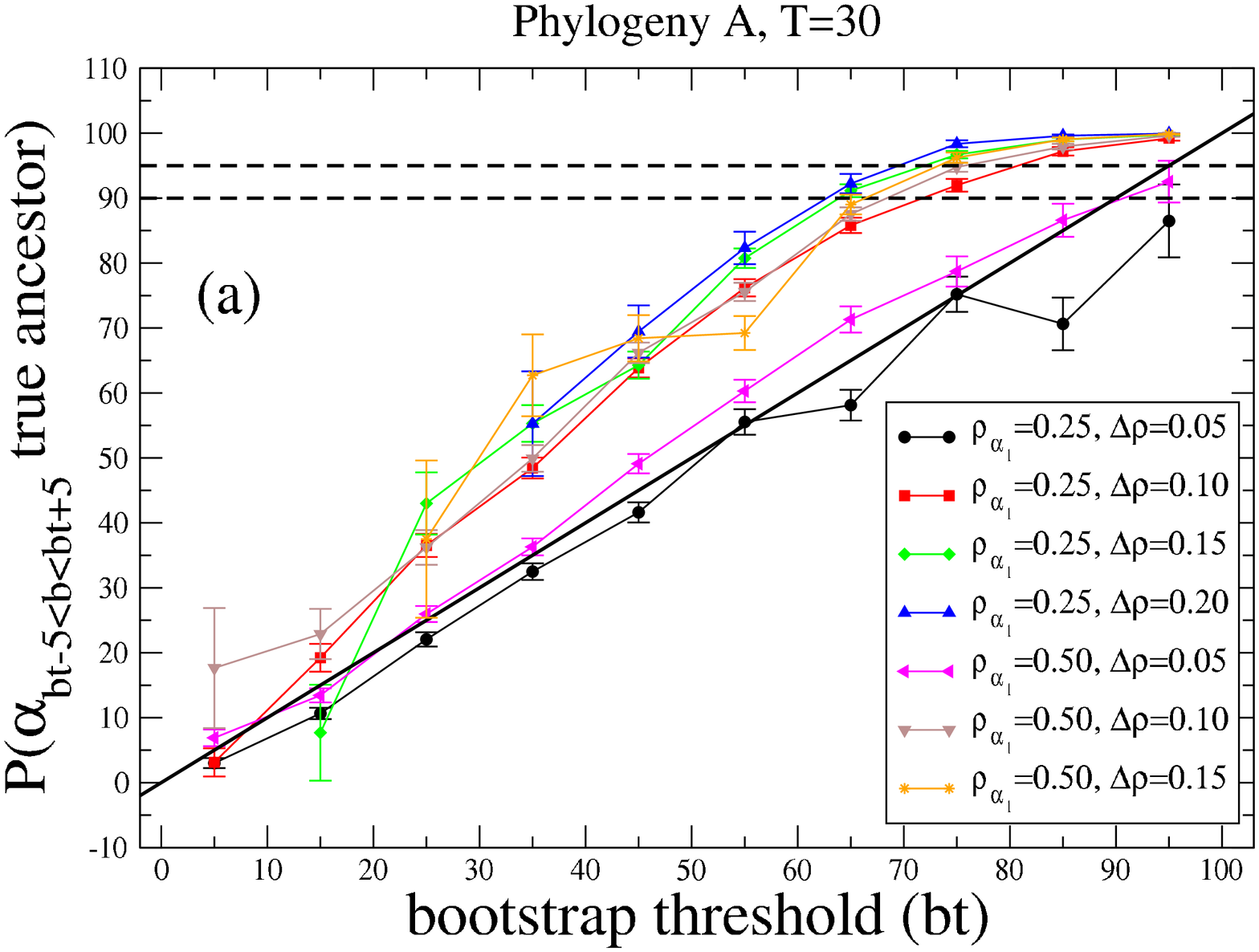}} 
             \resizebox{0.8\columnwidth}{!}{\includegraphics{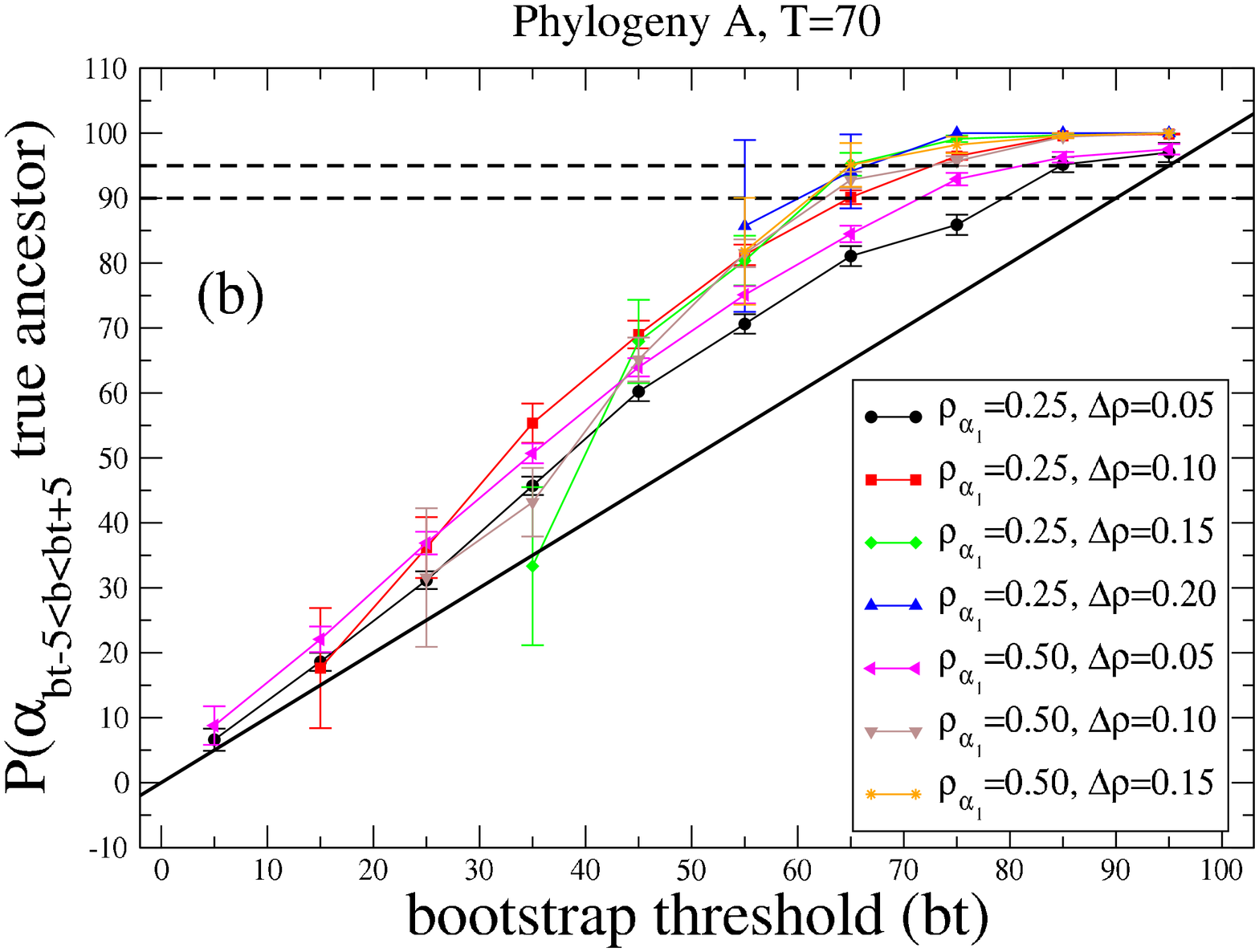}}
              \resizebox{0.8\columnwidth}{!}{\includegraphics{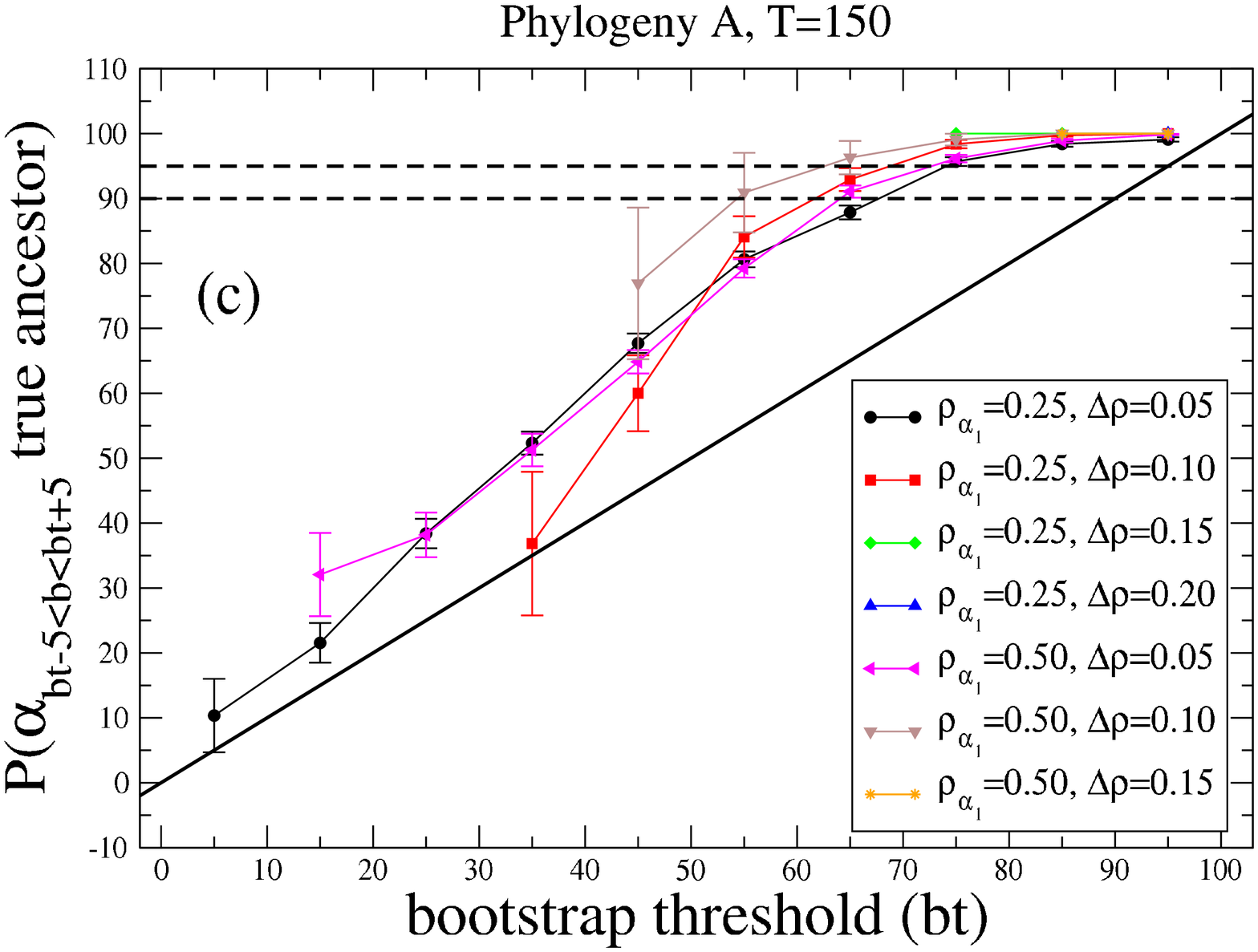}} 
              \caption{Probability that a node with bootstrap value in the range $[bt-5\%,bt+5\%[$ belongs to the phylogeny A. In the $x$ axis we report the bootstrap value as a percentage, whereas in the $y$ axis we report the discussed probability (as a percentage). The results shown in the figure are based on $S=1000$ simulations of series length $T=30$ in panel (a), $T=70$ in panel (b) and $T=150$ in panel (c), all of the simulations being performed by starting from the phylogeny A as discussed in the text and reported in Fig. \ref{DendroA}. The root node has been disregarded everywhere in the figure. The values of node correlations are also summarized in Table \ref{tab1}. Error bars in the figure correspond to one standard deviation estimated according to the binomial distribution. 
%The diagonal straight line indicates direct correspondence of $x$ and $y$ axes. 
%Results obtained for bootstrap values in a range that appeared less than 5 times over the 1000 simulations, i.e. less than the $0.07\%$ of the total number of nodes present in the simulations, are not shown in the figure.
}
\label{dendroAsimul}[b]
\end{center} 
\end{figure}

\begin{figure} 
\begin{center}
              \resizebox{0.8\columnwidth}{!}{\includegraphics{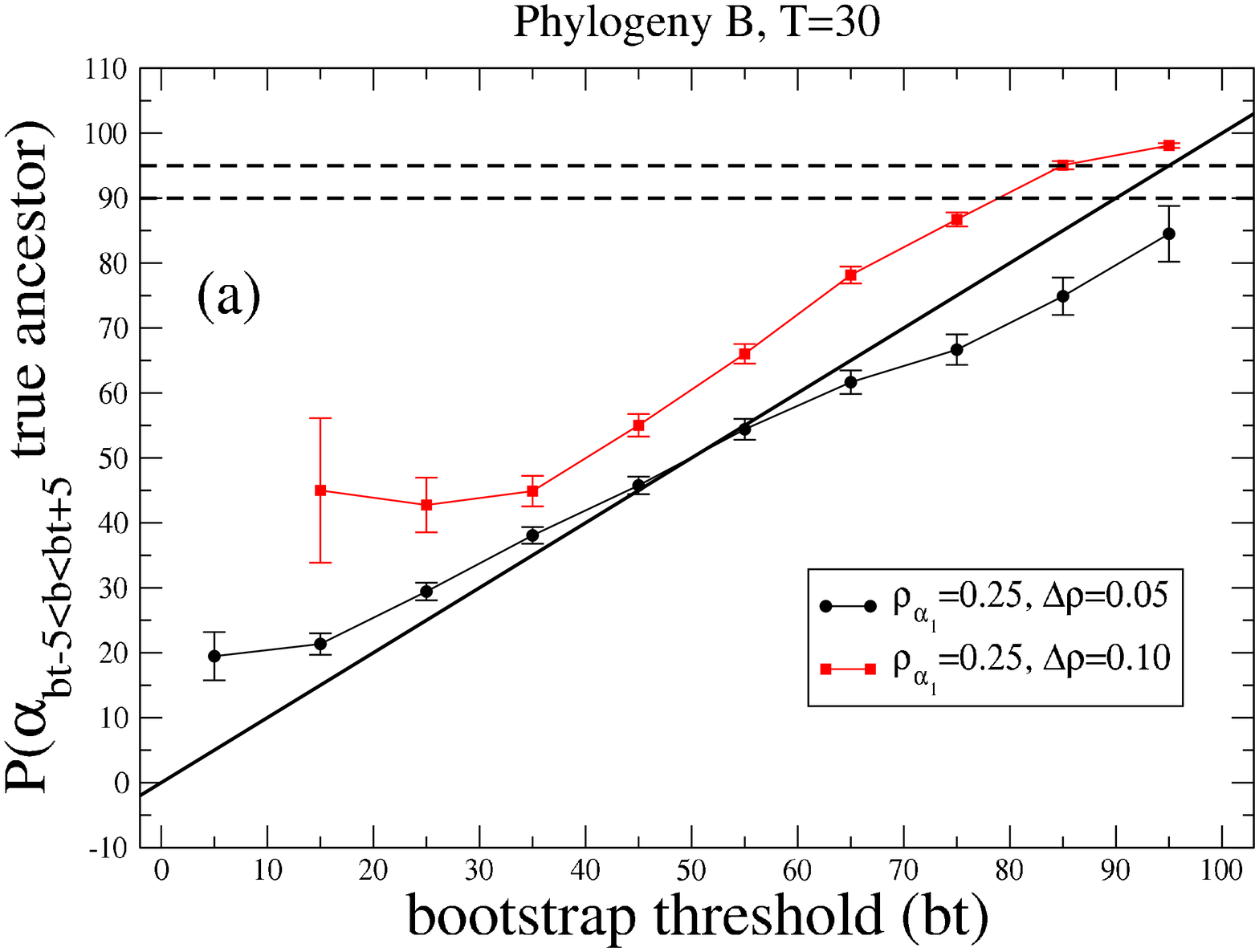}} 
             \resizebox{0.8\columnwidth}{!}{\includegraphics{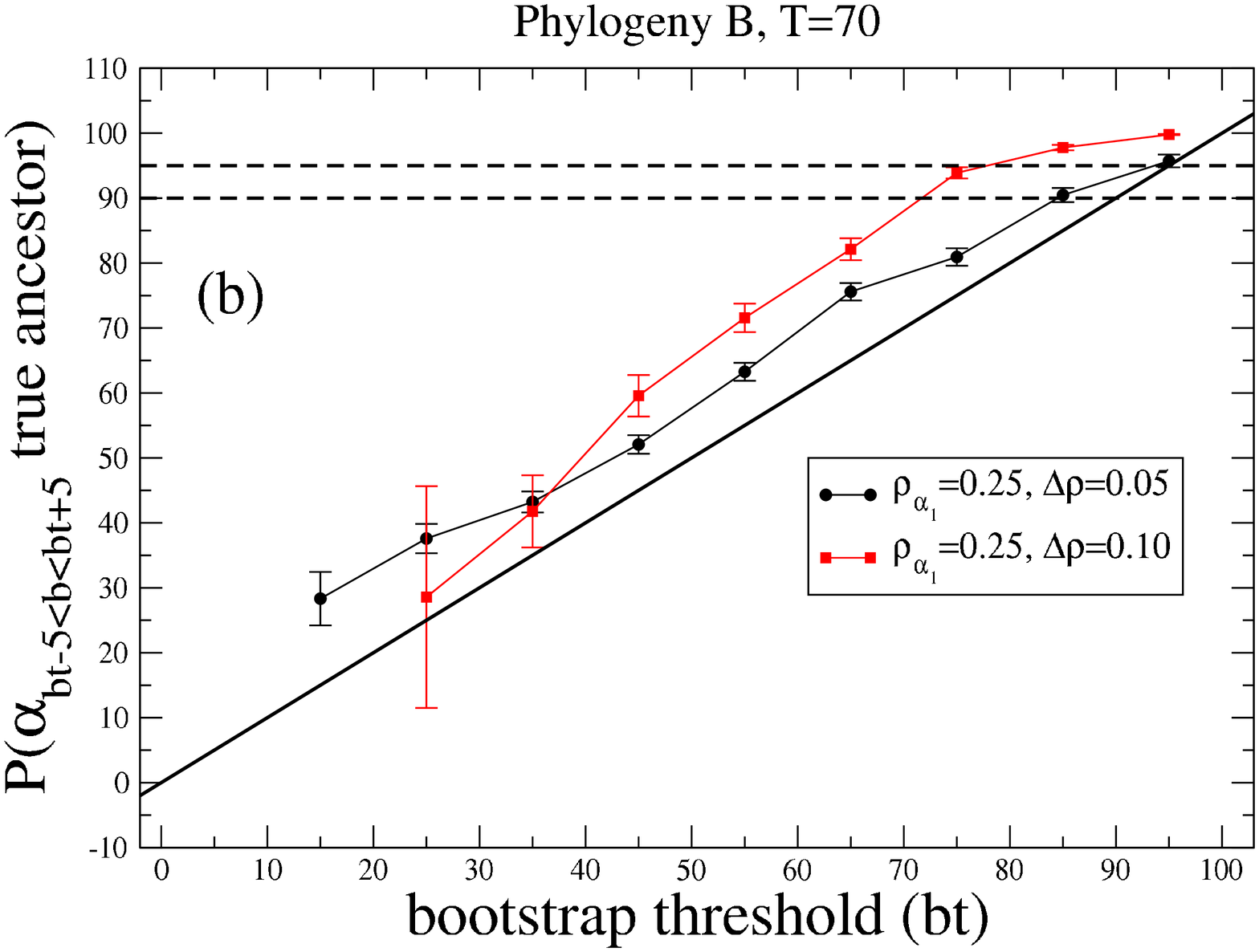}}
              \resizebox{0.8\columnwidth}{!}{\includegraphics{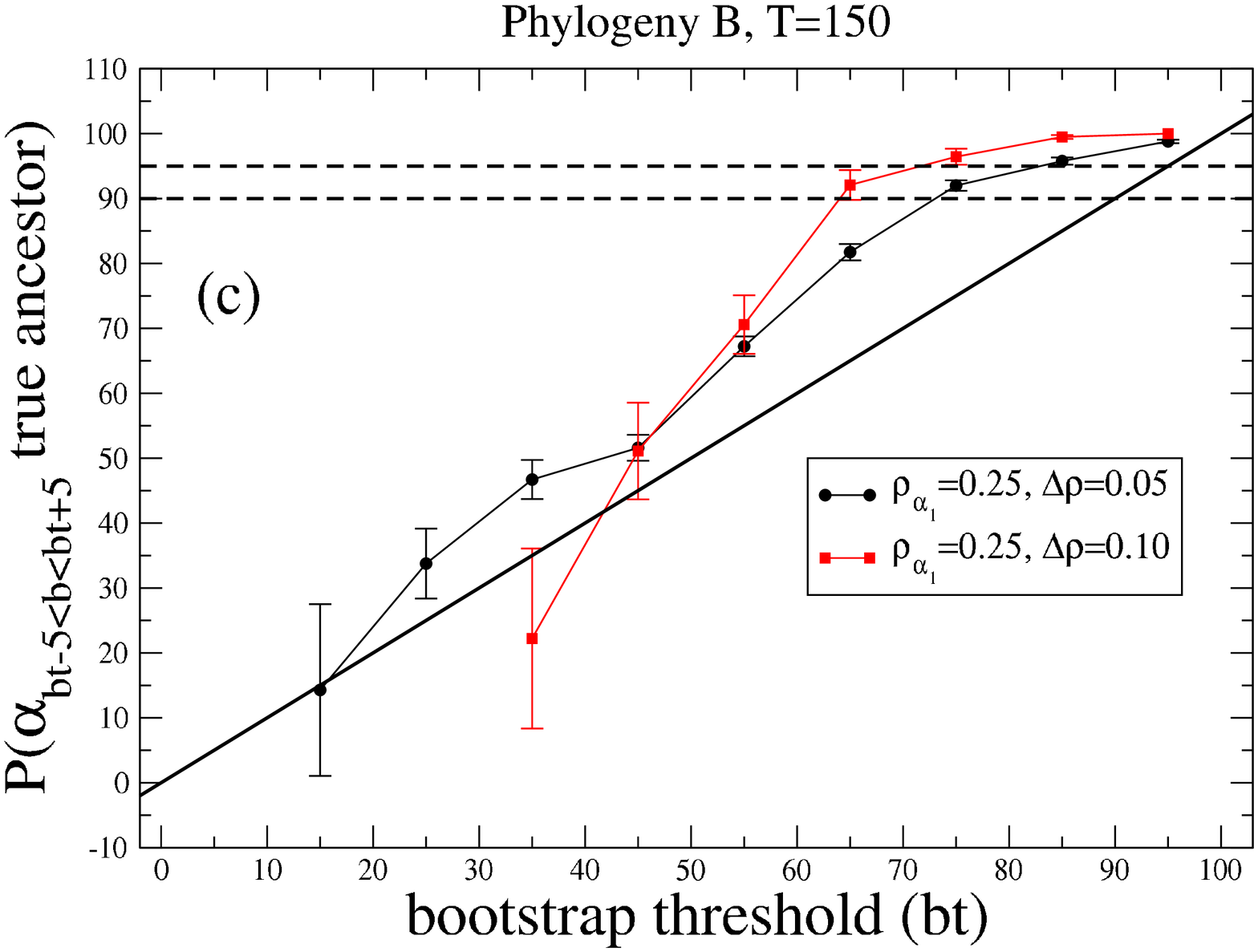}} 
              \caption{Probability that a node with bootstrap value in the range $[bt-5\%,bt+5\%[$ belongs to the phylogeny B of Fig. \ref{DendroB}. All of the simulations have been performed by starting from the phylogeny B. See the caption of Fig.\ref{dendroAsimul} for further details.} 
\label{dendroBsimul} 
\end{center} 
\end{figure}

Several parameters are involved in our investigation. Specifically, we set (i) the number $S$ of simulations of a given phylogeny and the number $B$ of bootstrap replicas that we have constructed for each simulation (we have set $S=1000$ and $B=100$) (ii) parameters describing the metric properties of the true phylogenies, i.e. the correlation value of nodes (see Table \ref{tab1}), and (iii) the length $T$ of symbol series.   
Fig. \ref{dendroAsimul} shows the results obtained for simulations based on phylogeny A, for three different data series length $T=30,70$ and $150$, each one corresponding to a specific panel in the figure. In each panel, we show the curves corresponding to all parameters reported in Table \ref{tab1}. Results obtained for bootstrap values in a range that appeared less than 5 times over the 1000 simulations, i.e. less than the $0.07\%$ of the total number of nodes present in the simulations, are not shown in the figure. The results reported in the figure indicate that, on average, the bootstrap value underestimates the probability of finding a node obtained from sample data to belong to the true phylogeny. Specifically, a node with a bootstrap value larger than $80\%$ usually corresponds to a true clade with a probability larger than $95\%$. These results are qualitatively similar to those obtained in Ref. \cite{Hillis}. It is however to notice that such behavior is not observed when both the length of data series is short ($T=30$) and $\Delta \rho=\rho_{\alpha_k}-\rho_{g(\alpha_k)}$ is sufficiently small, e.g. $\Delta \rho=0.05$ (see panel (a) of Fig. \ref{dendroAsimul}). It is also to observe that the curves are not sensibly affected by the absolute level of correlation $\rho_{\alpha_k}$, while the shape of the curve depends significantly on the relative correlation between two linked internal nodes, i.e. the branch length $\Delta \rho$. This suggests a sort of invariance for translation in the space of correlations. As an example of such a behavior we can look at panel (a) of Fig. \ref{dendroAsimul}, in which the curve corresponding to $\rho_{\alpha_1}=0.50$ and $\Delta \rho=0.05$ is much more similar to the curve corresponding to $\rho_{\alpha_1}=0.25$ and $\Delta \rho=0.05$ than, for instance, to the curve obtained for $\rho_{\alpha_1}=0.50$ and $\Delta \rho=0.10$. A similar behavior can also be observed in the other panels of the figure. By increasing the length of data series (moving from panel (a) to panel (c) of the figure) we note that curves tend to saturate at shorter values of bootstrap proportions. For instance, looking at panel (c) of Fig. \ref{dendroAsimul}, we note that a bootstrap value of $70\%$ is enough to get a probability larger than $95\%$ that the corresponding node belongs to the true phylogeny. Such a behavior is still more evident 
%if one compares results reported in Fig. \ref{dendroAsimul} and Fig. \ref{dendroBsimul} with results obtained 
for series of length $T=2000$. In this case even the most noisy configuration of correlations that we have considered here, i.e. $\rho_{\alpha_1}=0.25$ and $\Delta \rho=0.05$ produces very stable results. Specifically, 6984 of the total $(N-2) S=7000$ nodes analyzed in the simulations have a bootstrap value  larger than $90\%$ and each of these 6984 nodes corresponds to a correct clade in the original phylogeny. This result shows that for very long series the model exactly reproduces the true phylogeny.
Finally, a comparison of Fig. \ref{dendroAsimul} and Fig. \ref{dendroBsimul} shows that the topology of the phylogeny is not relevant in determining the relationship between bootstrap proportions and the probability of the corresponding clade being correct. 
%
%In summary our results shows that the relationship between bootstrap proportions and the probability of the corresponding clade being correct is sensitive to both the length $T$ of data series and the branch length $\Delta \rho$, whereas such a relationship is slightly affected by the topology of the true phylogeny and by the absolute level of correlation.

%\begin{widetext}
\begin{table}
\begin{center}
\begin{tabular}{||c|c|c|c|c|c|c||}
%{c|c h{1.97cm}h{1.97cm}h{1.97cm}}
%\colrule \colrule
\cline{1-3}
\bf{phylogeny} & $\rho_{\alpha_1}$ & $\Delta \rho=\rho_{\alpha_k}-\rho_{g(\alpha_k)}$\\
\cline{1-3}
\cline{1-3}
A & 0.25 & 0.05 \\
A & 0.25 & 0.10 \\
A & 0.25 & 0.15 \\
A & 0.25 & 0.20 \\
A & 0.50 & 0.05 \\
A & 0.50 & 0.10 \\
A & 0.50 & 0.15 \\
B & 0.25 & 0.05 \\
B & 0.25 & 0.10 \\
\cline{1-3}
%\colrule
%\colrule \colrule
\end{tabular}
\caption{\label{tab1} Setting list of the correlation value of nodes in the true phylogenies (A and B) that have been used in the simulations.}
\end{center}
\end{table}
%\begin{widetext}
%\end{widetext}

%\end{widetext}

\section{Conclusions}\label{conclusions}

In conclusion, we have introduced a general algorithm for generating multivariate symbolic sequences with a given hierarchical similarity structure. The fact that we do not make any assumptions on the generating mechanism for these sequences makes this algorithm useful in those cases when the dynamics generating the phylogeny is not known. 
We have used our algorithm in order to assess the bootstrap confidence in phylogenetic analysis. Our results show that, on average, the bootstrap value underestimates the probability of finding a node obtained from sample data to belong to the true phylogeny. This fact is qualitatively in agreement with the results obtained in Ref. \cite{Hillis}. However we have also observed that the relationship between the bootstrap proportion and the probability of the corresponding clade being correct is sensitive to both the length $T$ of data series and the branch length $\Delta \rho$, whereas such a relationship is only slightly affected by the topology of the true phylogeny and by the absolute level of correlation.

There are several extensions that could be made to our algorithm. First, as mentioned at the end of Section \ref{algorithm}, one can consider trees in which two leaves have different similarity with their closest common ancestor. This may be useful when one wants to model the possibility that the molecular clock is different in different branches of the tree. A second extension concerns the possibility of having models with correlations between different sites. In the current version of the model we have generated independently each site of the sequence. However it is known that different sites of DNA, proteins, etc., are in fact correlated. Our algorithm can be extended to reproduce dependencies between different sites.
Finally, our algorithm might be used to assess the role of the finite length of the series in discovering the true phylogeny. Imagine to have a set of short sequences and to ask how much the reconstructed phylogeny is affected by the sequence length. Our algorithm allows to generate sequences of arbitrary length, but preserving the similarity structure, and thus to answer the question.

\section{Appendix: Proof of Proposition I}

%Let us now suppose that an infinite number of rows ($T \rightarrow \infty$) has been generated. 
Consider two elements (leaves) of the system (dendrogram), say $i$ and $j$, merging first together at the node $\alpha_k$. What is the percentage of times in which the two elements took the symbol from the same node? Or, in other words, what is the probability $P(i,j)$ that the two elements take their symbol from the same node? Obviously the nodes involved are only those nodes connecting the node $\alpha_k$ to root node $\alpha_1$, both $\alpha_k$ and $\alpha_1$ included. In order to simplify the notation we indicate $g(\alpha_k)$ with $\beta_1$, $g(\beta_1)$ with $\beta_2$ and so on, following the path from the node $\alpha_k$ up to the root, i.e. $\alpha_1=\beta_q=g(\beta_{q-1})$. It results:
\begin{eqnarray}\label{probalphak}
P(i,j)=p(\alpha_k, \bar{\beta_1},\bar{\beta_2},...,\bar{\beta_q})+\nonumber \\
+p(\beta_1,\bar{\beta_2},...,\bar{\beta_q})+...+\nonumber \\
+p(\beta_{q-1},\bar{\beta_q}=\bar{\alpha_1})+p(\alpha_1)
\end{eqnarray}
where $p(\beta_t,\bar{\beta}_{t+1},...,\bar{\beta_q}=\bar{\alpha_1})$ is the joint probability that at a generic step of the protocol two leaves $i$ and $j$ take the symbol from the node $\beta_t$ and not from all $\beta_{t+1}$, $...$, $\beta_q=\alpha_1$. In order to show that the probability in Eq. (\ref{probalphak}) is equal to $\rho^I_{\alpha_k}$, we need to perform some intermediate calculations. The  probability that elements $i$ and $j$ do not take the symbol of the node $\beta_s$ conditioned by the fact that they didn't take the symbol from the nodes $\beta_{s+1}$, $...$, $\beta_q=\alpha_1$ is
\begin{eqnarray}\label{pnegcondneg}
p(\bar{\beta_s}|\bar{\beta}_{s+1},...,\bar{\beta_q}=\bar{\alpha_1})=\nonumber \\
=1-p(\beta_s |\bar{\beta}_{s+1},...,\bar{\beta_q}=\bar{\alpha_1})= \nonumber \\
%={\text because of Eq. (\ref{condnode}) }
=1-\frac{\rho^I_{\beta_s}-\rho^I_{\beta_{s+1}}}{1-\rho^I_{\beta_{s+1}}}=\frac{1-\rho^I_{\beta_s}}{1-\rho^I_{\beta_{s+1}}}.
\end{eqnarray}
where we have used the relation given in Eq.(\ref{condnode}).
Another relation that we need to state, in order to show that the probability in Eq. (\ref{probalphak}) is equal to $\rho^I_{\alpha_k}$, is
\begin{eqnarray}\label{pnegandneg}
p(\bar{\beta}_s,\bar{\beta}_{s+1},...,\bar{\beta}_q=\bar{\alpha}_1)=\nonumber \\
=p(\bar{\beta}_s |\bar{\beta}_{s+1},...,\bar{\beta}_q) \cdot p(\bar{\beta}_{s+1},...,\bar{\beta}_q)= \nonumber \\
%=\text{ because of Eq. (\ref{pnegcondneg}) }
=\frac{1-\rho^I_{\beta_s}}{1-\rho^I_{\beta_{s+1}}} \cdot p(\bar{\beta}_{s+1},...,\bar{\beta}_q)=\nonumber \\
=\frac{1-\rho^I_{\beta_s}}{1-\rho^I_{\beta_{s+1}}} \cdot \frac{1-\rho^I_{\beta_{s+1}}}{1-\rho^I_{\beta_{s+2}}}
\cdot ... \cdot \frac{1-\rho^I_{\beta_{q-1}}}{1-\rho^I_{\beta_{q}}} p(\bar{\beta}_q)=\nonumber \\
=\frac{1-\rho^I_{\beta_s}}{1-\rho^I_{\beta_{q}}} \cdot (1-\rho^I_{\beta_q})=1-\rho^I_{\beta_s}.
\end{eqnarray}
A generic term of Eq. (\ref{probalphak}) can therefore be written as
\begin{eqnarray}\label{pdirandneg}
p(\beta_s,\bar{\beta}_{s+1},...,\bar{\beta}_q=\bar{\alpha}_1)=\nonumber \\
=p(\beta_s |\bar{\beta}_{s+1},...,\bar{\beta}_q) \cdot p(\bar{\beta}_{s+1},...,\bar{\beta}_q)= \nonumber \\
%=\text{ because of Eq. (\ref{pnegcondneg}) }
=\frac{\rho^I_{\beta_s}-\rho^I_{\beta_{s+1}}}{1-\rho^I_{\beta_{s+1}}} \cdot p(\bar{\beta}_{s+1},...,\bar{\beta}_q)=\nonumber \\
%= \text{ because of Eq. (\ref{pnegandneg}) }
=\frac{\rho^I_{\beta_s}-\rho^I_{\beta_{s+1}}}{1-\rho^I_{\beta_{s+1}}} \cdot (1-\rho^I_{\beta_{s+1}})=\rho^I_{\beta_s}-\rho^I_{\beta_{s+1}}.
\end{eqnarray}
By introducing the result (\ref{pdirandneg}) into Eq. (\ref{probalphak}) and taking into account that $p(\alpha_1)=\rho^I_{\alpha_1}$ according to Step 2 of the protocol, we obtain
\begin{eqnarray}\label{probalphakSEC}
P(i,j)=\rho^I_{\alpha_k}-\rho^I_{\beta_1}+(\rho^I_{\beta_1}-\rho^I_{\beta_2})+...\nonumber \\
+(\rho^I_{\beta_{q-1}}-\rho^I_{\alpha_1})+\rho^I_{\alpha_1}=\rho^I_{\alpha_k}.
\end{eqnarray}
This equation shows that the probability $P(i,j)$ that two elements (leaves) $i$ and $j$, which merge together in the dendrogram at the node $\alpha_k$, take their symbol from the same node, is equal to $\rho^I_{\alpha_k}$. 

\bigskip

\noindent {\bf Acknowledgments }\\
\indent Authors acknowledge partial support from the European Union STREP project n. 012911 ``Human behavior through dynamics of complex social networks: an interdisciplinary approach".
% ----------------------------------------------------------------

\end{document}